\documentstyle[12pt,aaspp4]{article}
 
\lefthead{Secker and Harris}
\righthead{Dwarf Galaxies in Coma. I.}

\begin{document}

\title{Dwarf Galaxies in the Coma Cluster. I. Detection, 
Measurement and Classification Techniques.}

\author{Jeff Secker\\
Program in Astronomy, Washington State University,
Pullman, WA 99164-3113 USA\\
{\em secker@delta.math.wsu.edu}}
 
\author{William E. Harris\altaffilmark{1}\\
Department of Physics and Astronomy, McMaster University,
Hamilton, Ontario L8S 4M1 Canada\\
{\em harris@physics.mcmaster.ca}}
 
\altaffiltext{1}{Visiting Astronomer, KPNO, operated by AURA, Inc.\ under 
contract to the National Science Foundation.} 

\begin{abstract}
Deep $B-$ and $R-$band CCD images of the central $\sim 700$ arcmin$^2$
of the Coma cluster core have been used to measure the dwarf-galaxy
population in Coma.  In this paper, we describe a newly developed code
for automated detection, photometry and
classification of faint objects of arbitrary shape and size on digital
images.  Intensity-weighted moments are used to compute the positions,
radial structures, ellipticities, and integrated magnitudes of
detected objects.  We demonstrate that Kron-type $2 r_1$ aperture
aperture magnitudes and surface brightnesses are well suited to
faint-galaxy photometry of the type described here.  Discrimination
between starlike and extended (galaxy) objects is performed
interactively through parameter-space culling in several possible
parameters, including the radial moments, surface brightness, and
integrated color versus magnitude.  Our code is tested and
characterized with artificial CCD images of star and galaxy fields; it
is demonstrated to be accurate, robust and versatile.  Using these
analysis techniques, we detect a large population of dE galaxies in
the Coma cluster core.  These dEs stand out as a tight sequence in the
$R$, $($$B-R$$)$ color-magnitude diagram.
\end{abstract}


\section{INTRODUCTION} 

The Coma cluster of galaxies (A1656) is a rich, spherically symmetric,
and centrally concentrated galaxy cluster, and with a mean recession
velocity of $v \simeq 7200$ km/s (Colless \& Dunn 1996), it is the
nearest of the very rich (class 2 or more) Abell clusters.  In terms
of luminosity, the cluster core is dominated by two supergiants, the
elliptical NGC 4889 and the cD NGC 4874.  As in all rich clusters, the
dwarf elliptical (dE) and nucleated dwarf elliptical (dE,N) galaxies
are by far the most numerous galaxies in the dense environment of the
cluster core (Dressler 1980; Ferguson \& Sandage 1991; Ferguson \&
Binggeli 1994; Secker \& Harris 1996), though their contribution to
the total cluster {\it mass} is not significant (Bernstein et
al. 1995). It is these dEs which are the target of the study we
present here.  With $cz$(Coma) = 7200 km/s, 1 arcmin $= 20.94h^{-1}$
kpc.  For the purposes of this discussion, we adopt a distance scale
of $H_0 = 75$ km/s/Mpc, which yields a distance modulus $(R-M_{R}) =
34.91$.

Deep optical CCD images of rich galaxy clusters present a formidable
challenge to photometric measurement techniques.  The thousands of
detectable objects on such images span a huge parameter space in their
measured properties: image {\it morphology} must be divided into
starlike versus nonstellar objects, with the latter subdivided into
morphological classes for each of the giant cluster galaxies, dwarf
galaxies, and background galaxies.  The range in {\it luminosity} of
the cluster galaxies is typically at least 10 magnitudes, and their
{\it surface brightness} ranges from those of dense nuclei or globular
clusters down to the extreme low surface brightnesses of the faintest
dwarfs, often less than 1\% of the night sky.  Lastly, typical {\it
apparent diameters} range over almost two orders of magnitude, for
example (in our Coma images) from two to three pixels for unresolved
(starlike) objects, to $\sim 3-20$ pixels for dEs, to 60-100 pixels
for the largest cluster members.  In addition, a rich cluster field
generates spatial differences in CCD noise levels, spatial resolution,
field crowding, and global light gradients.  The data analysis problem
in its entirety is sufficiently challenging that any single technique
will be strained to measure all types of objects correctly.

Here we present a newly constructed code which is necessarily
restricted to one aspect of this overall problem, specifically {\em a
semi-automated approach to the detection, photometry and
classification of faint objects} on CCD images.  Specifically, we
describe techniques we have assembled to measure accurate total
magnitudes, integrated colors, and several intensity-weighted
quantities to discriminate among different image morphologies.  When
these techniques are used in conjunction with other pre-existing tools
(e.g., profile-fitting photometry for crowded fields, isophote-fitting
tools for modeling large-diameter giant cluster galaxies, and
median-filtering techniques to remove the slowly varying light
gradients), we have a sophisticated set of computer codes, sufficient
for a thorough analysis of CCD images of rich galaxy clusters.

In the following sections we describe the calculation of total
magnitudes, colors and intensity-weighted moments for faint objects on
large-format CCD images.  We have implemented these in a robust
software code which we call {\sc DYNAMO}; it has been rigorously
tested and characterized on artificial and real CCD data sets. The
{\sc DYNAMO} source code is well documented and it is available
electronically from the first author.

This is Paper I of two papers concerning dE galaxies in the Coma
cluster (Paper II is Secker, Harris \& Plummer 1997).  In Section 2 we
describe the observations and image preprocessing for our Coma cluster
images.  In Section 3 we discuss the estimation of accurate mean sky
values and object detection.  In Section 4 we discuss the computation
of unbiased intensity-weighted quantities, together with their use in
image analysis and classification. In Section 5 we describe the
calculation of total magnitudes, integrated colors and surface
brightnesses, and show preliminary results for the Coma cluster dE
sample.

\section{Coma Cluster CCD Observations}

On the nights April 9 and 10, 1991, we used a TE $2048^2$ CCD at the
prime focus of the KPNO 4-m telescope to mosaic-image the Coma cluster
core.  Exposure series in both the $R$- and $B$-bands were obtained
for three overlapping core fields as well as a control field placed
well outside the cluster, as listed in \ref{TAB1paper1}.  Provided in
Table \ref{TAB1paper1} are the field centers as specified in the image
headers, and in Table \ref{TAB2paper1} we provide a summary of
parameters relevant to these images.  In addition to the program and
control fields, throughout both nights numerous images of standard
star fields were obtained (Table \ref{TAB3paper1}), which were used
for calibration.

\placetable{TAB1paper1}
\placetable{TAB2paper1}
\placetable{TAB3paper1}

Gray-scale images of the three program fields are shown in Figure
\ref{COMAfields}. 
The {\sc NGC 4874} field is roughly centered on the cD galaxy NGC
4874, and it overlaps (by about 400 pixels) with the {\sc NGC 4889}
field to the East and the {\sc NGC 4874 South} field to the South. The
{\sc NGC 4889} field includes the supergiant galaxy NGC 4889, and
extends outwards to the east.  The {\sc NGC 4874 South} field is
located below NGC 4874, with the southernmost edge extending $\simeq
23.33$ arcmin, or $\simeq 490h^{-1}$ kpc, from NGC 4874.  The {\sc
Control} field is located a full 2.09 degrees, or $2.63h^{-1}$ Mpc,
almost directly east of the cluster center at NGC 4874.

\placefigure{COMAfields}

The TE2K CCD had 27 micron pixels, an image scale of $0^{''}.53$/px, a
gain of 21.4 $e^-$/{\sc ADU} and a readnoise of 13 $e^{-}$/px.  During
the two (photometric) nights of observation, the seeing stayed
consistently in the range of 1.1 to 1.3 arcsec.  This particular CCD
was an early-generation large-format detector and suffered from rather
poor cosmetics, dominated by the completely unusable columns 1701
through 2044. When trimmed, the field of view was about
15$^{\prime}$$\times$18$^{\prime}$.  With this and the overlapping
regions taken into account, we observed a total area of $\simeq 700$
arcmin$^2$, a significant fraction of the Coma cluster core.  The
trimmed {\sc Control} field has an area of 270 arcmin$^2$.  While our
original intention was to image two additional fields south of NGC
4874 and another control field, midway through the second night the
CCD detector suffered electronic failure, terminating the data
acquisition.

\subsection{Image Preanalysis Processing}

Image preprocessing was carried out within IRAF\footnote{{\sc IRAF} is
distributed by the National Optical Astronomy Observatories, which are
operated by the Association of Universities for Research in Astronomy,
Inc., under contract to the National Science Foundation.}, with the
usual routines {\em noao.imred.ccdred.ccdproc} and {\em
images.imcombine}.  The raw images were first overscan corrected,
trimmed and bias subtracted.  Flat fields were constructed from a
combination of dome flats, twilight flats, and dark-sky flats (from
the sparsely-populated {\sc Control} field).  The individual exposures
were averaged together to yield {\em eight master images}; $R$- and
$B$-band master images for each of the three program fields and the
control field. As specified in Table \ref{TAB2paper1}, two of the
$B$-band master images represent a median of the three exposures,
while the other two $B$-band master images represent a a straight
average of the two available images.  The $R$-band master images are
all weighted averages of three individual 900-sec exposures, where the
weighting factors are given by

\begin{equation}
w_i = \frac{1}{\sum_{i=1}^{n} w_i} \left( \frac{I_i}{z_{s_i}^2 
\sigma_i^2} \right).
\end{equation}

\noindent Here, $z_{s_i}^2$, $\sigma_i^2$ and $I_i$ represent the mean sky
level in the vicinity of the star, the star's {\sc FWHM} and a measure
of its sky-subtracted central intensity.  Thus higher weights are
assigned to those exposures with lower sky values and sharper seeing
profiles.  While the S/N ratio of the $B$-band images is lower than
for the corresponding $R$-band images at equivalent magnitudes we used
$B$ only for color index measurement and not for object detection.

The $B$-band master images of each field were aligned with the
$R$-band master images, with the flux-conserving {\sc IRAF} routines
{\em images.geomap} and {\em images.geotran}.  At the same time, the
transformations between program fields were derived to map both the
{\sc NGC 4889} field and the {\sc NGC 4874 South} field to the
coordinates of the {\sc NGC 4874} field.  In this system, the centers
of NGC 4874 and NGC 4889 are located at approximately (1061,1010) and
(1875,1124), and the bright (saturated) star above NGC 4874 is located
near (1000,1735).

In order for our measured aperture colors to be accurate, we must
ensure that the sky-subtracted intensities summed within the
individual apertures are independent of differences in seeing between
the $B$- and $R$-band master images. Since the $B$ images had slightly
better seeing quality than the $R$ images, we convolved the $R$ master
frames with a Gaussian filter of width $\sigma_G = \sqrt{\sigma_{\rm
B}^2 -\sigma_{\rm R}^2}$.  Numerical values for $\sigma_G$ are given in
Table \ref{TAB2paper1}.  The end result is that for all four fields,
the seeing ({\sc FWHM}) of the $R$-band master images agrees with the
$B$-band images to an accuracy of about $\Delta${\sc FWHM}$\simeq$0.01
pixels. At this level of accuracy, we can be sure that our color
estimates are consistent and independent of seeing effects.

All eight of the master images were then filtered wth the fast ring
median filter of Secker (1995) to remove all small-scale structure; 
the ring-filtered versions of the master
images were then subtracted from the originals, and the appropriate
sky levels added back in, yielding a final image with all large-scale
light gradients removed to facilitate faint-object detection and
photometry.  The characteristic radius $r_c$ of the ring filter should
be chosen as small as possible for the population of interest:  we
used 16 weighted pixels (see Secker 1995) in an annulus of characteristic
radius $r_c = 16$ pixels ($2.96h^{-1}$ kpc at Coma).  Visual
inspection shows that this radius is appropriate for all but the 
$\sim 100$ brightest galaxies (giants plus bright dwarfs) on each of the
program fields. These brightest galaxies were measured on the
original (unfiltered) master images to ensure that their true
luminosity was not underestimated. (Because of the light gradients
present on the unfiltered images, the consequent uncertaintes 
in their total magnitudes are larger.)

\subsection{Artificial CCD Images}

To facilitate the testing and characterization of our photometry and
classification techniques, we used {\sc IRAF}'s artificial
image/object generation packages ({\em noao.artdata.gallist,
.starlist, .mkobject}) to generate artificial CCD images closely
resembling the features of the real data (intensity profiles,
sky-pixel scatter, range in apparent magnitude and object diameter).
The artificial CCD images are, however, cosmetically perfect,
uncrowded, and have no overall light gradients.  The results of these
simulations are described at the appropriate positions throughout this
text.  In general, we find that our intensity-weighted moment and
total magnitude estimates are systematically quite accurate.

Two $1024^2$ simulated images were generated, each with an equivalent
gain of 64.2 $e^-$/{\sc ADU}, a read noise of 7.51 $e^-$/px and a mean
background level of 1500.0 {\sc ADU}, together with the corresponding
Poisson noise.  To each image one distinct population of object
(either stars or galaxies) was added.  While any selection of galaxies
and stars which span a reasonable magnitude will enable us to test our
code, the input luminosity functions of the stars and galaxies were
matched to those observed for the Coma cluster. These luminosity
functions are defined over $15.0 \leq R \leq 24.0$ mag.  This range
includes the brightest dE galaxies, and extends fainter than our $R =
22.5$ limiting magnitude (see below and Paper II). The spatial
distribution of all objects was determined by random sampling from a
uniform distribution; in this manner, crowding effects are minimized.

The first image consisted of 500 Gaussian-profile stars, with a {\sc
FWHM} = 3 px, corresponding to Gaussian standard deviation of
$\sigma_{\rm G} = 1.27$ pixels.  The input luminosity function
followed a power law $n(m) \sim 10^{0.4m}$.  The second image
consisted of 500 exponential-profile galaxies convolved with the
seeing profile, with minor to major axis ratios of $0.1 \leq b/a \leq
1.0$.  Our choice of an exponential profile is consistent with current
observations for the majority of dE galaxies (Ferguson \& Binggeli
1994).  The input luminosity function followed a Schechter function
with a characteristic magnitude of $R^{\star}=15.50$ mag and a
power-law exponent of $\alpha = -1.45$.

\section{Sky Estimation for Object Detection and Photometry}

For all but the most crowded fields, the set of pixels in a CCD image
is dominated by those which do not contain flux from resolved objects
such as stars or galaxies.  In most broadband images these {\em sky
pixels} have a mean value, and a scatter about this mean, governed
primarily by Poisson statistics.  For typical images of galaxies in
the field or in cluster environments, the images are only sparsely
covered by objects, such that the number of {\em object pixels} is
$\lesssim 5$ percent.  Thus, the probability distribution function of
pixel intensity values is generally dominated by a Gaussian
distribution for the sky pixels, skewed by a bright-side tail
extending to high intensity values (Figure \ref{PIXELhist}).  While
the contamination by number is small, the contamination arising from
the large intensity values of these object pixels is extreme, and
therefore the estimates of a mean sky value are not trivial.  In our
approach to faint galaxy photometry, there are two separate instances
in which we depend upon accurate and consistent estimates of the
average sky intensity: a {\it global} estimate to set an object
detection threshold for the entire CCD image, and {\it local}
estimates from annuli about each object.  For both the global and the
local estimates, the sample of pixels will be contaminated by object
pixels.  Thus, we implement an iterative method to estimate the {\em
mode} and {\em standard deviation} for a vector of pixel intensity
values, using Chauvenet's criterion (Parratt 1961) to trim obviously
discrepant (non-sky) pixel intensities.  This routine is based on
techniques developed for {\sc DAOPHOT II} (Stetson et al. 1990;
Stetson 1994), and it consistently matches the results of Stetson's
code (to within a scatter of $\pm 0.2$ {\sc ADU}) when applied to
similar data vectors.

\placefigure{PIXELhist}

\subsection{Global Sky Estimates and Object Detection}

For any given CCD image, the faint and low-surface-brightness objects
are most difficult to detect. One technique to increase the detection
efficiency for faint {\em stellar} objects is to pass a convolution
kernel over the CCD images prior to utilizing a detection algorithm
(e.g. Stetson 1994).  This technique works well for stars, which have
a well defined and nearly uniform radial intensity profile across the
field, but not for galaxies, each of which has a distinct intensity
profile.  We therefore adopt a different algorithm for object
detection.  Provided the primary CCD image is globally
flat\footnote{This can be accomplished, for example, using a
combination of isophotal modeling to subtract the bright galaxies, and
a median filter (Secker 1995) to remove large-scale light gradients
(cf. Section 2.2.)}, we assume simply that each local intensity
maximum exceeding some constant intensity threshold is associated with
the center of a unique object.  In our analysis of the Coma CCD images
we perform object detection on the (deeper) $R$-band master images,
which we refer to as {\em primary} images.  The $B$-band images are
referred to as {\em secondary} images, used only to construct $(B-R)$
color indices for the primary sample of objects.

To set the global sky level on the median-filtered master images, we
initially select $4\times10^4$ unique and random pixels covering the
full area of the image.  The subset of these pixel values which are
within the valid intensity range form a data vector from which the
global mode $\overline{z_{\rm sky}}$ and standard deviation
$\overline{\sigma_{\rm sky}}$ are iteratively estimated.  We then
define the detection threshold (Figure
\ref{PIXELhist}) as an intensity level, some multiple $\beta$ of
$\overline{\sigma_{\rm sky}}$ above $\overline{z_{\rm sky}}$.  For an
object to be detected it must have at least one pixel which exceeds
this threshold.  The user-specified dimensionless parameter $\beta$ is
typically in the range 3.5-5.0, with the lower limit set to avoid
detecting bright peaks in the Poisson noise distribution.  Pixel
values falling below the detection threshold can not (by our
definition) be designated as object centers.

All pixels on the $R$-band master images satisfying $z_i \geq \left(
\overline{z_{\rm sky}} + \beta
\overline{\sigma_{\rm sky}} \right)$ are potential {\em object} pixels, 
and these are grouped for further analysis.  These object pixels
compose a data vector which is sorted by intensity into descending
order, and then divided into two groups: (1) pixels defining the
peak-flux intensity for potential object centers, and (2) those pixels
which are {\it connected neighbors} of a potential object center
(i.e. they are immediately adjacent to an existing object). We do this
separation by defining the location of the first (brightest) pixel as
the peak-flux position of the first object.  For the remaining
objects, we take advantage of the fact that for individual objects,
the pixel intensity values decrease rather smoothly as the radial
distance from the center increases (at least until the Poisson noise
in the object's wings begins to dominate the signal).  Thus stepping
through the ensemble of all object pixels, each subsequent pixel will
form an element of the first object at increasing radii, or will be
the peak-flux pixel of a new object.  We defined a logical function
called {\em neighbor}, which we use to divide the pixels into their
two groups; it returns {\em false} if the $i^{\rm th}$ pixel is
isolated (i.e., separated from an existing peak-flux object center by
lower-intensity pixel values; i.e., group 1), and returns {\em true}
if this $i^{\rm th}$ pixel is separated from a peak-flux object center
by less than the seeing disk radius, or if the pixel is attached to a
brighter pixel of the same object, forming part of the decreasing
continuum of pixels belonging to that object (i.e., group 2).  By
applying the {\em neighbor} function to the sorted vector of pixels
brighter than the detection threshold, we obtain our list of center
positions (integer peak-flux pixel coordinates $x_{\rm cen}$ and
$y_{\rm cen}$) for individual objects.

For our Coma data, we adopted a detection threshold of $\beta = 4.0$,
which corresponds to $\simeq 1.5 - 2$ percent of the mean sky level.
Although objects with a central surface brightness lower than this
level are {\em measurable}, attempts to {\em detect} these faintest
objects objectively also yield many spurious detections from pure
noise.  Our method (like most others) is therefore strongly biased
against extremely low surface brightness objects.  We do not regard
this as a serious problem in our case, as these extreme-LSB objects
will mostly be fainter than the adopted limiting magnitude $R = 22.5$
mag (Paper II).

On each master $R$-band image, we detected on the order of
$4\times10^3$ objects to this central surface brightness threshold. Of
these objects, at least 50 percent are fainter than $R = 22.5$ and
were discarded.  In addition, we visually inspected {\it every}
detected object, and deleted obvious ``false'' detections,
corresponding to CCD defects, charge overflow columns and diffraction
spikes around bright stars.

\subsection{Local Sky Estimates For Aperture Photometry}

The estimate of an object's magnitude is essentially an exercise in
counting. We count the total number of photons received within the
area of an object, and subtract from this the number of photons which
would have been counted even if the object were not there (i.e., the
{\em local} mean sky level). But this is not a trivial task, and for
faint objects uncertainty in the local sky level is the limiting
factor in our ability to compute accurate and unbiased photometry
(cf. Section 5.5).  To compute accurate values for object total
magnitudes, one must first estimate a mean sky value (in {\sc ADU}) at
the exact location of the object in question (starlike or
non-stellar).  However, since the object and sky share the same set of
pixels, the determinations of the total magnitude and local sky value
are coupled.

We choose to compute accurate local sky values by iteratively
estimating the intensity mode in an annulus about each object.  The
inner radius of the annulus is chosen large enough to minimize
contamination from the object, yet small enough that the estimate is
representative of the local sky.  The outer radius is set to ensure a
that a statistically large number of pixels are included in the
estimate.  $R_{\rm in}$ (pixels) denotes the inner radius of this sky
annulus, and $W_{\rm ann}$ (pixels) denotes the width of the sky
annulus.  While $R_{\rm in}$ and $W_{\rm ann}$ would remain constant
for stars (all of which have the same profile on a given image), they
will differ among non-stellar objects. Thus for accurate
photometry of faint galaxies, $R_{\rm in}$ and $W_{\rm ann}$ must be
chosen with care to reflect the projected dimensions of the galaxies
of interest.

For each detected object, the radial distance of the $i^{\rm th}$
pixel from the peak-flux pixel is represented by $r_i$, and a data
vector of intensity values is constructed from all pixels which
satisfy the criteria $R_{\rm in} \leq r_i \leq R_{\rm in}+W_{\rm
ann}$.  The resulting data vector is used to compute the local sky
mode $z_{\rm sky}$ and the local standard deviation $\sigma_{\rm
sky}$.  These estimates are made for every detected object, and they
are used to derive the object's magnitude and color, as described in
Section 5.

We separated the detected objects into three sets, dependent
upon their apparent diameters. These groups are (i) the brightest and
most extended giant galaxies, (ii) other giants and the brightest dEs,
and (iii) the numerous faint and small-scale galaxies.  Photometry for
objects in the latter group is measured on the filtered-subtracted
master images, with a sky annulus of inner radius 24 pixels and a
width of 4 pixels.  Objects in first two groups are measured on the
original master images with the sky annuli chosen appropriately: an
inner radius of 95 pixels for group (i) and 45 pixels for group (ii),
and a width of 4 pixels in both cases.

\section{Image Parameter Analysis}

\subsection{The Limiting Radius For Summations}

In this paper, the measurement and classification of faint objects on
CCD images is based on the computation of accurate intensity-weighted
moments.  For low signal-to-noise (S/N) data, these calculations
depend strongly on the estimated sky value in the vicinity of
individual objects, and the radial distance over which the moment
summations are carried.  For the radial moments discussed below,
there is an inherent difficulty: in order to carry the
intensity-weighted summations to a sufficiently large radius, one
needs to know the true extent of an object.  However, it is the radial
moments themselves which define the object's size.  If the summations
are prematurely truncated, the resultant moments will be
underestimated.  If these sums are carried too far in radius, the
resultant moments can be rendered meaningless, since the
sky noise will dominate the intensity contribution to the sums.

We therefore implement a curve-of-growth technique to estimate the
limiting radius $R_{\ell}$ for the calculation of the
intensity-weighted radial moments.  We define $R_{\ell}$ to be the
radius at which the object's flux drops below one percent of the local
sky value.  Thus $R_{\ell}$ defines the radius of a circular aperture
interior to which the computation of the radial moments is carried
out.  This radius is sufficiently large to enclose the object, yet it
it is not strongly affected by contributions from sky pixels 
to the intensity-weighted summation.  As will be seen below,
$R_{\ell}$ is used only in the calculation of
the image moments and directly for any magnitude calculations.

The limiting radius $R_{\ell}$ is computed separately for each object
as follows.  Concentric circular annuli are placed around the
peak-flux pixel (i.e., $x_{\rm cen}$ and $y_{\rm cen}$), and the {\em
mean} pixel intensity values within these annuli are computed.
Denoting the interior and exterior edges of each annulus by $r_{\rm
in}$ and $r_{\rm out}$, the smallest annulus has $r_{\rm in} = 2.5$
pixels, each annulus has a constant width of two pixels, $r_{\rm out}
= r_{\rm in} + 2$, and with each concentric annulus $r_{\rm in}$ is
incremented by 0.5 pixels.  For each annulus, the
mean pixel intensity $<$$z$$>$ is computed for the set of pixels
satisfying

\begin{equation}
r_{\rm in}^2 \leq [(x_i-x_{\rm cen})^2+(y_i-y_{\rm cen})^2]
< r_{\rm out}^2; \ \ \ \ \ \ i=1,...,n.
\end{equation}

For stars and early-type galaxies, this mean intensity will typically
decrease with radius outwards, approaching the local sky level.  If
the local sky mode $z_{\rm sky}$ is the true sky level near the object
in question, $R_{\ell}$ is defined as the radius at which the mean
pixel intensity first satisfies $<$$z$$>$$ \leq 1.01 z_{\rm sky}$,
where $R_{\ell}$ is the geometric mean of $r_{\rm in}$ and $r_{\rm
out}$ ($R_{\ell} = \sqrt{r_{\rm in}r_{\rm out}}$) or the radius of the
seeing disk, whichever is greater.  In this manner $R_{\ell}$
represents the radius at which the object's signal merges (at the one
percent level) with the local sky level.  Thus $R_{\ell}$ will enclose
a similar fraction of the total flux for all exponential-profile
galaxies.  While this fraction will differ for stars and de
Vaucouleurs-$R^{1/4}$ profile galaxies, $R_{\ell}$ itself is not used
directly for photometry (Section 5.1).

The summations involved in the calculation of intensity-weighted
radial moments (Section 4.3) include pixels interior to $R_{\ell}$;
thus, if $R_{\ell}$ is systematically underestimated, the
intensity-weighted moments will also be.  The results presented in
Section 4.3 show that this is not the case.  However, when 
calculating $R_{\ell}$ a source of error arises from nearby objects, 
which can cause an overestimate for $R_{\ell}$.  We have implemented a
facility to enable rapid verification of each object on an individual
basis: simple, low-resolution intensity plots are generated, centered
on each object, with the calculated $2r_1$ radius clearly
displayed. For objects affected by nearby neighbors, $R_{\ell}$ can be
modified interactively, with the radial moments recalculated interior
to this new value, minimizing the effect of nearby objects.

\subsection{Intensity-Weighted Averages For Object Centers}

The object detection routine described in Section 3.1 provides object
centers in terms of $x_{\rm cen}$ and $y_{\rm cen}$, which represent
the (integer) position of the peak-flux pixel.  However,
intensity-weighted averages in the vicinity of the peak-flux pixel
yield center positions accurate at the fractional pixel level.  The
center positions corresponding to the intensity-weighted averages are
denoted $<$$x$$>$ and $<$$y$$>$, and are given by

\begin{eqnarray}
<x> & = & \frac{\sum xI}{\sum I}  \nonumber \\
<y> & = & \frac{\sum yI}{\sum I}.
\label{eq:centers}
\end{eqnarray}

\noindent  Above, $x$ and $y$ represent image pixels, 
and $I$ represents the sky-subtracted pixel intensities, given
by $I = z_i - z_{\rm sky}$.

The summations in (\ref{eq:centers}) are computed for all pixels
interior to an aperture of radius equal to two pixels.
Adopting this small (and constant) radius yields sufficient
pixels to obtain accurate weighted moments, yet it is small enough to
constrain the center estimate to be nearby the peak-flux pixel center.
The center positions calculated for objects on the artificial CCD
images compare well with the input values.  The estimates are
unbiased, and the scatter is on the order of $\pm$0.2 pixels for
bright stars and galaxies, increasing to $\pm$0.4 pixels for the
faintest objects.

\subsection{The Intensity-Weighted Radial Moments}

The intensity-weighted radial moments $r_1$ and $r_{-2}$ were defined
in their integral form by Kron (1980), and in a discrete form by
Harris et al. (1991).  The first moment, $r_1$, and the second
negative moment, $r_{-2}$, of the radial intensity distribution, are
given by

\begin{eqnarray}
r_1    & = & \frac{\sum r I}{\sum I} \nonumber \\
r_{-2} & = & \left( \frac{\sum I/(r^2+0.5)}{\sum I} \right)^{-1/2}.
\label{eq:r1}
\end{eqnarray}

\noindent These two radial moments have units of pixels, and are 
constrained to lie in the range $(0,R_{\ell})$.  Above, $r$ represents
the distance (in pixels) of the $i^{\rm th}$ pixel from the object's
peak-flux pixel center, $I$ is the sky-subtracted intensity value, and
the summations are taken over the ensemble of pixels which satisfy
$r^2 \leq R_{\ell}^2$.  In (\ref{eq:r1}), the arbitrary additive
constant of 0.5 is included to avoid singularities, while the 
--1/2 exponent is included to return the dimensions of length.

These two radial moments in principle provide different and
complementary descriptions of the radial intensity distribution for
individual objects, since the $r_1$ moment is an indicator of the
image wing spread and thus of physical dimension (see below), whereas
the $r_{-2}$ moment quantifies the degree of central concentration of
an object. For both moments, starlike objects have a similar low value
constant for all magnitudes (i.e., the stellar sequence), and
non-stellar (galaxy) images scatter (to larger values) away from the
stellar sequence. Since the $r_{-2}$ moment gives highest weight to
the brightest central pixels, it is stable and robust, and in general
less affected by crowding than the $r_1$ moment.  In practice, these
two moments are also well correlated for nonstellar objects (Harris et
al. 1991).  Analysis of our artificial galaxy field CCD image shows
that the resolved galaxies are reasonably described by $r_1 \simeq
2.83 r_{-2} - 2.43$, with both moments measured in pixels.

These two radial moments have been calculated for all objects on both
of the artificial CCD images, and when plotted against total magnitude
(Figure \ref{RDLMNTsim}), the difference between stellar and
nonstellar objects is immediately obvious.  The radial moments for the
stellar objects are $r_1 = 1.6$ and $r_{-2} = 1.4$ pixels over the
entire magnitude range and provide a valuable consistency test that
the limiting radius $R_{\ell}$ encloses almost all the luminosity.
For bright stars, the majority of the luminosity is clearly measured,
and the first radial moment is constant at its theoretically-predicted
value of $r_1 = \sqrt{\pi/2} \sigma_{\rm G}$.  For faint stars, the
$r_1$ and $r_{-2}$ moments scatter about this true value, yet the
average value remains remarkably constant.  While we do not calculate
formal uncertainties for these radial moments, their scatter about the
stellar sequence provides a reasonable estimate of their uncertainty
as a function of magnitude.

\placefigure{RDLMNTsim}

Parameter space plots of $r_{-2}$ versus magnitude have been effective
in culling nonstellar objects from samples of starlike objects (e.g.,
Harris et al. 1991; Butterworth \& Harris 1992; Fleming et al. 1995;
see also McLaughlin et al. 1995 for a variant of this approach), and
it is this technique which we adopt here for our star/galaxy
discrimination.  The scattering of non-stellar objects away from the
constant stellar sequence is evident in Figure \ref{RDLMNTsim}.  Also
evident is the magnitude (in this case, near $R \simeq 21$ mag) at
which the nonstellar sequence merges with the starlike sequence.  This
is a direct result of the small linear size of the faintest galaxies
coupled with the CCD image scale and seeing.

In Figure \ref{FOURrdlmmnts}, we plot the intensity-weighted radial
moment $r_{-2}$ versus the apparent $R_{2r_1}$ magnitude, for objects
detected on the four $R$-band CCD images described in Section 2.1.
Numerous nonstellar objects are present at all magnitudes; near
$R_{2r_1} \simeq 19.5$ mag, these begin to merge with
the stellar sequence, and below $R_{2r_1} \simeq 21.5$ mag, as the
nonstellar objects become faint and starlike, the $r_{-2}$ moment
loses its ability to discriminate, and all objects appear starlike.
Evident on all four plots in Figure \ref{FOURrdlmmnts} are two
additional trends. First, the bright end of the stellar sequence rises
slightly above the typical mean value, most probably caused by a
change in image structure due to the behavior of the CCD near
saturation.  Second, a population of objects appears which are faint
and pointlike; i.e., their calculated $r_{-2}$ moment is less than for
a stellar profile.  These arise from cosmic rays, hot pixels or CCD
defects.

\placefigure{FOURrdlmmnts}

While there are nonstellar objects evident in all four of the CCD
fields, as expected, the {\sc Control} field has considerably fewer
nonstellar objects.  The dotted lines in Figure \ref{FOURrdlmmnts} are
used to restrict the data sample.  The vertical line at $R_{2r_1} =
22.5$ mag corresponds to our magnitude limit, and all objects fainter
than this are discarded. The upper horizontal line excludes objects in
and below the bright stellar sequence, with $r_{-2} \le 1.6$ and
$R_{2r_1} \le 19.5$ mag.  The number of starlike objects in this
region are 63, 59, 67 and 64 for the fields (a) through (d) in Figure
\ref{FOURrdlmmnts}. Within Poisson statistics, these numbers are
entirely consistent with each other.  Finally, all objects located
below the lower horizontal line near $r_{-2} \simeq 1.25$ (i.e., the
pointlike artifacts) are discarded.

The faint end of the stellar sequence deserves further comment. 
While the fractional contribution of galaxies in the range $R_{2r_1}
\le 19.5$ mag and $r_{-2} \le 1.6$ is negligible, for $R_{2r_1} >
19.5$ mag the starlike sequence is also populated with compact and/or
unresolved galaxies.  These galaxies are more numerous on the program
fields (within the cluster), and discarding the complete stellar
sequence would impose an extreme bias against unresolved galaxies.
Thus for these, we simply use the total control field number density
to define the background level of unresolved objects.

\subsection{The Scale Radius For Galaxies}

In 1980, Kron illustrated that the $r_1$ moment provides a measure of
the half-light radius for galaxies.  Here, we illustrate that the
$r_1$ moment is directly related to the scale radius $r_0$ of
exponential-profile objects (e.g., dE galaxies).  An exponential
profile, defined in terms of the central surface brightness $I_{\rm
c}$, with $r$ measured along the semi-major axis, is given by

\begin{equation}
I(r) = I_{\rm c} \exp (-r/r_0).
\label{eq:exppro}
\end{equation}

\noindent Evaluating the integral form of (\ref{eq:r1}) with $I(r)$
given by (\ref{eq:exppro}) yields $r_1 = 2 r_0$. In practice, the
measured value of $r_1$ is slightly underestimated (since the integral
is evaluated out to $r = R_{\ell}$ and not to $r = \infty$), and the
true radii for the smallest and faintest galaxies are essentially
unrecoverable without much better image resolution.  The {\it
half-flux radius} $r_{1/2}$ provides a different measure of the
physical dimension of galaxies.  The flux interior to a specific
radius $F(r)$, for the exponential profile given by (\ref{eq:exppro}),
yields

\begin{equation}
\frac{F(r)}{F(\infty)} = 1 - (1+r/r_0)\exp (r/r_0).
\label{eq:flux}
\end{equation}

\noindent The half-flux radius is obtained by setting (\ref{eq:flux}) 
equal to 1/2; we then find $r_0 = 0.596 r_{1/2}$.

\subsection{Ellipticity Estimation Via Moment Calculation}

To measure the ellipticity of faint objects, we adopt a method
involving the second-order intensity-weighted central moments
described by Valdes, Tyson \& Jarvis (1983; henceforth VTJ), given
by
 
\begin{eqnarray}
M_{\rm xx} & = & \frac{\sum (x-x_{\rm cen})^2I}{\sum I}  \nonumber \\
M_{\rm yy} & = & \frac{\sum (y-y_{\rm cen})^2 I}{\sum I} \nonumber \\
M_{\rm xy} & = & \frac{\sum (x-x_{\rm cen})(y-y_{\rm cen})I }{\sum I},
\end{eqnarray}
 
\noindent with the summations taken over all pixels satisfying $r_i \leq
R_{\ell}$, and with $I = z_i - z_{\rm sky}$.  VTJ define two 
quantities involving $a$ and $b$ (the semi-major and semi-minor axis 
lengths), given by

\begin{eqnarray}
(a^2 - b^2) & = & [(M_{\rm xx} - M_{\rm yy})^2 + (2 M_{\rm xy})^2 ]^{1/2} 
\nonumber \\ 
(a^2 + b^2) & = & M_{\rm xx} + M_{\rm yy}.
\label{eq:semiaxes}
\end{eqnarray}

\noindent Then, the ellipticity, $\epsilon$, and the eccentricity, $e$, 
are related in the normal manner by
 
\begin{equation}
\epsilon = 1 - \frac{b}{a} = 1 - \sqrt{1-e^2}.
\end{equation}
 
\noindent With the quantities calculated in (\ref{eq:semiaxes}), a definition
of the ellipticity (ours differs from that used by VJT) is given by
 
\begin{equation}
\epsilon = 1 - \left\{ \frac{(a^2 + b^2) - (a^2 - b^2)}{(a^2 + b^2) + 
(a^2 - b^2)} \right\}^{1/2}.
\end{equation}
 
\noindent   It is important to note that the ellipticity computed from 
the moments measured for individual objects will underestimate the
true ellipticity for elongated (and faint) objects.  This occurs
because corrections for small guiding errors and finite seeing are not
taken into account (refer to VTJ), and because the moments are
unstable (a very low signal-to-noise ratio) at these low flux levels.
Figure \ref{ELLIPcomp} illustrates how the ellipticity estimates
scatter about the true value for all objects on the artificial images.
The ellipticity is clearly underestimated for the complete range of
galaxy magnitudes, but especially for the fainter galaxies.  On the
other hand, $\epsilon$ is over estimated for the faint,
intrinsically-round, stellar images.  While the ellipticity estimates
for the brighter starlike and nonstellar objects are nearly unbiased,
for the faint dE galaxies $\epsilon$ is too uncertain
to prove useful for this analysis.

\placefigure{ELLIPcomp}

\section{Faint Galaxy Photometry}

The techniques currently in use for faint galaxy photometry include
fixed or variable circular apertures, isophotal boundaries, and
curve-of-growth analysis and profile-fitting methods (e.g., Kron 1980;
Jarvis \& Tyson 1981; Yee 1991; Lilly et al. 1991; Bershady et
al. 1994; De Propris et al. 1995; Fischer et al. 1997).  The approach
we use in the DYNAMO code is based on aperture photometry, where (as
indicated above) the aperture size depends on the scale radius of the
particular object.  If the CCD images are {\em uncrowded}, such that
the typical distance between galaxy centers exceeds twice their radii,
conventional aperture magnitude techniques provide an accurate method
for faint galaxy photometry.  Quite simply, aperture magnitudes count
the photons received by the detector; nothing more, and nothing less.
In addition, aperture techniques are straight-forward; magnitudes and
colors are simple to calculate and reproducible.

\subsection{Total Magnitude Estimation}

The definition of an object's total magnitude is straight forward to
state --  sum the sky-subtracted flux contained within an
aperture enclosing all of the object.  In practice, this is
non-trivial to implement for an automated galaxy survey. The method
which we adopt is to define an {\em individual aperture radius for
each object}, a radius which is proportional to a measure of its
physical dimensions.  Kron (1980) showed that the radial moment $r_1$
(Section 4.3) provides a good estimate of the half-light radius for an
ensemble of faint galaxy images.  Both Kron (1980) and Bershady et
al. (1994) define a circular aperture of radius of $2r_1$, which when
centered on the galaxy encloses the majority of its flux, even for
rather elongated objects.  They determine that the $2r_1$ radius
measures about 90 percent of the flux for exponential-profile
galaxies, and about 96 percent for stellar profiles; these are
easily corrected to true total magnitudes using simulations.  However,
the flux measured by the $2r_1$ radius is significantly less,
and thus underestimates the true total magnitude, for galaxies with de
Vaucouleurs  $R^{1/4}$ intensity profiles.  Thus, we must ask of
the ensemble of galaxies we are observing, which have exponential
profiles?  As discussed in Section 2.3, our targets of interest, Coma
cluster dEs, have an exponential profile.  And as discussed by
Trentham (1997), cluster dwarf irregular galaxies and the majority of
background galaxies in the range $20 < R < 25$ mag also have
exponential profiles.  This leaves the Coma cluster early-type giants.
As these ellipticals compose only a small fraction of our sample, we
simply note that the magnitude correction factor we derive below
(appropriate for dEs) may lead to an underestimate of the total
magnitudes for these ellipticals.

Bershady et al. (1994) provide a detailed discussion of the $2r_1$
total magnitude estimation technique.  Briefly, it provides a
compromise between total flux and low photometric error, and it is
accurate and robust.  We therefore adopt this $2r_1$ magnitude,
denoted by a subscript $2r_1$, as a measure of total magnitude,
realizing that a simple numerical factor will be necessary to obtain
true total magnitudes\footnote{Analysis of our artificial galaxy-field
CCD image illustrates that a radius of $3r_1$ measures virtually
unbiased total magnitudes for exponential-profile galaxies.  However,
the photometric error on the magnitude estimates is greatly increased
over the $2r_1$ magnitudes, and the larger radius poses definite
problems for objects in crowded fields.}.  Note that the radius used
for the $2r_1$ aperture is defined to be $2r_1$ {\em or the radius of
the seeing disk}, whichever is the greatest. For the $R$-band Coma
images, the seeing disk radius used was 3.5 px, which is a $3
\sigma_{G}$ radius enclosing 99.73 percent of the flux for stellar
objects.  We found, as did Bershady et al.  (1994), that the $2r_1$
magnitude is insensitive to changes in seeing, and the slight
differences in seeing between our four $R$-band master images do not
significantly affect our estimates of total magnitude.

Once $r_1$ has been calculated accurately, the total instrumental
magnitudes can be calculated, by integrating the total flux contained
within a circular aperture centered on the peak-flux pixel of the
object.  The method used here to calculate the total enclosed flux
considers fractional-pixel flux, which reduces the ragged-edge effect
apparent for small aperture radii (Stetson et al. 1990).  For the
ensemble of object pixels, $r_i$ denotes the radial distance of
the $i^{\rm th}$ pixel from the peak-flux center, and $I$ represents the
sky-subtracted pixel intensity.  Then the fraction of the $i^{\rm th}$
pixel which is interior to $2r_1$ is given 
by $F_i = {\rm max}\left[0.0,{\rm min}(1.0, 2r_1 - r_i+0.5) \right]$, 
and this fraction of the pixel's flux is included in the summation as
as $\sum I = \sum I + I_i F_i$.  The total area $A_{\rm ap}$ within 
$2r_1$ is given by the sum of $F_i$ for the set of object pixels.
Then the total instrumental magnitude is given by

\begin{equation}
M_{2r_1} = \overline{ZP} - 2.5 \log(\sum I/t_{\rm e}),
\label{eq:obmag}
\end{equation}

\noindent normalized to a one-second exposure by the exposure time
$t_{\rm e}$.  This is converted (roughly) to an apparent magnitude
with the magnitude zero-point $\overline{ZP}$, which represents the
magnitude of a star for which the CCD would measure 1 {\sc ADU} per
second.  While $\overline{ZP}$ is determined from stellar objects, it
represents a simple conversion factor and is directly applicable to
all nonstellar objects as well.

The $R_{2r_1}$ total magnitude was calculated for all objects on both
of the artificial CCD images. Figure \ref{RMAGcorrections} illustrates
that for both stars and galaxies, this $R_{2r_1}$ magnitude
consistently provides a reasonable measurement of the true total
magnitude, $R_T$, and for $R_T \lesssim 22.75$, the scatter in $\Delta
R$ remains less than 0.1 mag\footnote{The bright galaxies measured
$\sim 0.4$-mag too faint represent objects at the edge of the CCD
image, while the 10 or so objects measured $\sim0.1$-$0.4$-mag too
bright represent inherently faint objects superimposed on brighter
objects.}.  Ideally, Kron's magnitude would measure {\it constant
fraction} of the total light, independent of the galaxy's distance or
apparent diameter. However, our simulations reveal that the light
contained within this $2r_1$ aperture varies from about 90 percent for
bright galaxies, to 100 percent for faint galaxies.  This occurs as
the successively fainter galaxies make the transition from large and
resolved to small and unresolved, with all of their light then
contained within the seeing profile.

\placefigure{RMAGcorrections}

For galaxies with $R_T = \gtrsim 16$ mag, we adopt a correction factor
constant at $\Delta R = R_T - R_{2r_1} = -0.12$ mag, keeping in mind
that this is an underestimate for giant ellipticals with an $R^{1/4}$
profile.  For galaxies in the range $16 \lesssim R_T
\lesssim 24$ mag, the magnitude correction factor varies linearly
between $\Delta R \simeq -0.12$ and 0.0 mag, and we adopt a correction
factor given by

\begin{equation}
\Delta R = 0.015 R_{2r_1} - 0.36.
\label{eq:deltaR}
\end{equation}

\noindent Over the same magnitude interval, the correction factor for
stellar objects varies between $\simeq 0.04$ and 0.0. For our Coma
data set, we discard objects on the stellar sequence brighter than
$R_{2r_1} = 19.5$ mag.  Below this, we can not discriminate between
starlike and non-stellar objects, and we correct total magnitudes for
all objects using (\ref{eq:deltaR}).  Thus while our final sample of
objects consists of stars and galaxies, any differences between their
magnitude correction factors are negligible.

In summary, magnitudes and colors determined via aperture techniques
are robust and accurate provided the image is not crowded.  Dealing
with crowded objects is unavoidably a complex problem.  For crowded
starlike objects, the iterative least-squares simultaneous fitting of
multiple stellar profiles (e.g.  Stetson 1994) is an efficient and
accurate technique.  For crowded galaxy objects, deconvolution by
profile fitting is also possible (e.g., Fischer et al. 1997).  However, this
is a very different approach then our aperture photometry.  In this
analysis we use the relative number of objects separated by less than
6 pixels (i.e. where the seeing disks are overlapping) as a measure of
the degree of crowding.  For our Coma cluster fields, the number of
multiple objects was in the range $\simeq 2-4$ percent of the total,
even on the {\sc Control} field. Since this number is so low, this set
of objects was discarded, with the assumption that the statistical
correlations derived from the data set will not be affected
significantly.

\subsection{Measures of Surface Brightness}

We define two measures of surface brightness, the first of which is
the central surface brightness $I_{\rm c}$ (mag/arcsec$^2$), defined
in terms of the peak-flux pixel intensity.  Here, $\Delta S$
represents the pixel area (arcsec$^2$), and $I_1$, $t_{e}$ and
$\overline{ZP}$ represent the sky-subtracted intensity of the
peak-flux pixel, the exposure time and the magnitude zero-point.
Then, the central surface brightness (mag/arcsec$^2$) is given by

\begin{equation}
I_{\rm c} = \overline{ZP} - 2.5 \log10( I_1) + 2.5 \log10(\Delta S t_e).
\end{equation}

\noindent For our CCD images of the Coma cluster, $\Delta S = 0.53^2$ 
arcsec$^2$, and for galaxies in Coma, $I_{\rm c}$ measures the 
central $185h^{-1} \times 185h^{-1}$ pc$^2$.

In Figure \ref{SIMcentsurf} we plot $I_{\rm c}$ versus total magnitude
$R_{T}$ for the stars and galaxies on our artificial CCD images.  For
both stars and exponential-profile dE galaxies, a decrease in $I_{\rm
c}$ corresponds to a decrease in total magnitude, as expected.  For
stellar objects with their constant Gaussian profile, $I_{\rm c}$ is
linearly related to $R_{2r_1}$, with significant scatter in the
relationship beginning near $R_{2r_1} \simeq 22$ mag. At any
magnitude, the exponential-profile galaxies have a lower central
surface brightness. In fact, star/galaxy discrimination can be
performed rather well in this parameter space, with the advantage
that the calculation of $I_{\rm c}$ does not depend upon summations
sensitive to the limiting radius $R_{\ell}$.

\placefigure{SIMcentsurf}

Second, we define an average surface brightness, $SB_{2.5\sigma}$,
which considers pixels interior to an aperture of radius $2 r_1$.  Of
all pixels inside this circular aperture, only those $N_p$ pixels
which satisify $I_i \geq z_{\rm sky} + 2.5 \sigma_{\rm sky}$
contribute to $\sum I$, the sky-subtracted pixel sum within the
aperture.  Then the average surface brightness is given by

\begin{equation}
SB_{2.5\sigma} = \overline{ZP} - 2.5 \log10( \sum I) + 2.5 \log10(\Delta S N_p 
t_e).
\end{equation}

\noindent Note that $SB_{2.5\sigma}$ yields an accurate measure of an
object's average surface brightness provided that there are no other
objects within the $2r_1$ aperture radius.  As well, there will be a
small contribution to $\sum I$ from Poisson sky noise; in most
instances this will be negligible, since this noise scatters about the
mean to both higher and lower intensity values.

\subsection{Color Estimation and Color-Magnitude Diagram Analysis}

Much of the controversy over the accuracy of integrated colors boils
down to whether or not there are color gradients within dE and dE,N
galaxies.  If a uniform color gradient exists within the galaxy (i.e.,
if the $B$-band intensity profile falls off differently than the
$R$-band profile), then a constant radius enclosing some fraction of
the total $R$ luminosity will include a different fraction of the $B$
luminosity.  However, recent observations show no such systematic
effects in these galaxies.  Observations of dE galaxies in Fornax
(Caldwell \& Bothun 1986; Cellone, Forte \& Geisler 1994) reveal no
color gradients, and similar observations of Virgo dwarfs (Vader et
al.  1988; Durrell et al. 1997; each study examined about 10 dEs) show
evidence for color gradients in up to 50 percent of the dEs, yet both
red and blue gradients are observed.  As well, the luminous nucleus in
the bright dE,N galaxies in our sample pose no problems, since the
colors of the nuclei are not noticeably different from their
surroundings (Caldwell \& Bothun 1986; Durrell et al. 1997; Vader et
al. 1988).

We choose to measure colors within an aperture constant for all
objects and on all images, with a constant radius comparable to the
{\sc FWHM} of a stellar profile.  Since the S/N ratio within a small
aperture of radius $R_{\rm ap} = 3$ (px) is greater than for the full
$2r_1$ aperture, this fixed aperture technique yields very accurate
colors, provided that the seeing is the same in both filters (cf.,
Section 2.3).  In addition, the fixed aperture colors are independent
of the calculation of the $r_1$ radius, and for small $R_{\rm ap}$
they are robust in the presence of nearby objects.  At the peak-flux
pixel position of the $i^{\rm th}$ object on the $R$ and $B$ images,
the local sky mode is calculated, and the sky-subtracted luminosity is
summed within a circular aperture of fixed radius $R_{\rm ap} = 3$
pixels ($0.56h^{-1}$ kpc for the distance of Coma).  The pixel
selection and luminosity summations within the circular aperture are
performed in the same manner as those described in Section 5.1.  Then
the $(B-R)$ color is defined as the difference of aperture magnitudes,
and the corresponding uncertainty is given by summing the individual
$R$- and $B$-band aperture magnitude uncertainties in quadrature
(cf. Section 5.5).  Note that for any reasonable detection threshold,
many of the objects for which we measure colors would not be detected
by themselves on the $B$-band image.

In Figure \ref{FOURcmds} we plot the CMDs for objects measured on the
three cluster fields and on the {\sc Control} field.  The dE galaxies
are immediately obvious as an excess of objects on the cluster fields,
over and above those objects on the {\sc Control} field.  The dE
galaxies are confined to a well-defined sequence of objects which
begins at $R_{2r_1} = 15.5$ mag near $(B-R) = 1.6$ mag and extends down to
the limiting magnitude at $R_{2r_1} = 22.5$ mag, spreading in width
due to the photometric errors in both magnitude and color (cf. Figure
\ref{PHOTerrors}).   Comparing the four CMDs in Figure \ref{FOURcmds}, 
it is clear that the dE galaxy sequence is continuous, that it is a
continuation of the sequence defined by the brighter cluster giants,
and that its faint end merges with the faint noncluster galaxies and with
Galactic G and K main-sequence dwarf and post-main sequence
giant stars.  The location of the dE sequence in the CMD is consistent
with colors found by Thompson \& Gregory (1993) for Coma cluster dEs,
and with  integrated colors of globular star clusters (Hopp et
al. 1995).  Our analysis of the mean dE galaxy color distribution
is presented in Paper II.

\placefigure{FOURcmds}

These CMDs provide a convenient parameter space to cull our sample of
objects and remove the obvious noncluster galaxies.  The vertical
dotted lines at $(B-R)=0.7$ mag and at $(B-R)=1.9$ mag represent
generous limits to the dE galaxy sequence.  Outside of this color
range, the majority of the objects are foreground Galactic stars and
galaxies more distant than the Coma cluster: for $(B-R)
\lesssim 0.7$ mag, all four CMDs are very sparsely populated,
and for $(B-R) \gtrsim 1.9$, the density of objects on the cluster
field CMDs is similar to the {\sc Control} field.  The lower
horizontal dotted line at $R_{2r_1} = 22.5$ mag corresponds to our
adopted magnitude and color completeness limit (Paper II); at this
magnitude limit we are $\gtrsim 80$ percent complete in detection and
almost 100 percent complete in color.  All objects fainter than this
limit are discarded.  Thus, in Paper II we restrict our analysis to
the subset of the objects within the color range $0.7 < (B-R) <1.9$
mag and $R_{2r_1} = 19.5$ mag, greatly decreasing the contamination in
our sample of candidate dE galaxies.  In addition, we use the
distribution of objects in the {\sc Control}-field CMD to
statistically correct the sample of objects detected on the program
fields, to further reduce the effects of contamination.

Lastly, we note that fainter than $R_{2r_1} = 22.5$
mag, there appears to be an increase in the number of blue objects on
each of the three program field CMDs, over and above the number on the
{\sc Control} field.  This excess is consistent with a spread in the
measured dE colors which increase dramatically past 
$R_{2r_1} \simeq 23$ mag.  A similar spread does not occur on the red
side because of the strong $B-$band incompleteness there.

\subsection{Uncertainties For Aperture Magnitudes}

Uncertainties on our $2r_1$ magnitudes and our fixed-aperture colors
are estimated in the manner described by Harris et al. (1991).  We
begin by estimating the uncertainty on sky-subtracted intensity sum
$\sum I$, for which the total variance results from three significant
contributions.  We denote the area of the sky annulus (integer pixels)
used to estimate $z_{\rm sky}$ by $A_{\rm sky}$, the standard
deviation of the pixel-to-pixel scatter (in {\sc ADU}) in the sky
annulus by $\sigma_{\rm sky}$, and the number of pixels in the
aperture containing the object by $A_{\rm ap}$.  First, the variance
due to photon noise from the object itself
is given by $\sum I g$ ($e^-$), where $g$ is the
CCD gain ($e^-$/{\sc ADU}).  Second, the random scatter in the sky
pixels under the object contributes a variance equal to $A_{\rm ap}
\sigma_{\rm sky}^2 g^2$ ($e^-$), which corresponds to product of the
number of pixels in the aperture and the variance per pixel.  Third,
the standard error of the mean for the local sky intensity is given by
$\sigma_{\rm sky} (A_{\rm sky})^{-1/2}$ ($e^-$), so that the
variance arising from a possible systematic error in the modal sky
level is given by $\left( A_{\rm ap} \sigma_{\rm sky} g \right)^2
A_{\rm sky}$.

The standard deviation in $\sum I$ (in adu) is given by

\begin{equation}
e(\sum I)_i  = \left[  \frac{\sum I}{g} + A_{\rm ap} \sigma_{\rm sky}^2
\left(1+\frac{A_{\rm ap}}{A_{\rm sky}} \right) \right]^{1/2}.
\label{eq:intuncer}
\end{equation}

\noindent For faint stars and for normal situations where $A_{\rm ap}
\ll A_{\rm sky}$, the Poisson sky noise
dominates, and (\ref{eq:intuncer}) reduces to $e(\sum I)_i \simeq
\sigma_{\rm sky} \sqrt{A_{\rm ap}}$.  
If we define upper and lower
limits on $\sum I$ to be $I_{\rm max} = \sum I + e(\sum I)_i$ and
$I_{\rm min} = {\rm max}(\sum I - e(\sum I)_i)$, 
then a rough error estimate on the
magnitude is half of the magnitude difference, given by

\begin{equation}
\sigma(M_{\rm ap}) = 0.5 \left[ 2.5 \log \frac{I_{\rm max}}{I_{\rm min}} 
\right].
\label{eq:maguncer}
\end{equation}

Then, the total uncertainty on our aperture magnitudes and colors is
given by $\sigma(M_{\rm ap})$ added in quadrature with the uncertainty
in the magnitude zeropoint, $\sigma(ZP)$, derived from standard star
observations.  In Figure \ref{PHOTerrors} we plot the uncertainties in
$R_{2r_1}$ magnitude (dots) and color (crosses) versus $R_{2r_1}$
magnitude for all detected and measured objects on our four progam
fields.  In summary, typical uncertainties for $R_{2r_1} < 22.5$ mag
are $\lesssim \pm0.06$ mag in total magnitude, and $\lesssim \pm$0.12
mag in color.

\placefigure{PHOTerrors}

\acknowledgments

This paper is based upon thesis research conducted by J.S. while at
McMaster University.  The research was supported in part by: the
Natural Sciences and Engineering Research Council of Canada (through a
grant to W.E.H.), the Department of Physics and Astronomy at McMaster
University, a grant to J.S. from NASA administered by the American
Astronomical Society, a Fullam Award to J.S. from Dudley Observatory,
and the Ontario Ministry of Colleges and Universities (through an
Ontario Graduate Scholarship to J.S).  We would like to thank
P.R. Durrell, P. Fischer, S. Lilly, D.E. McLaughlin, C.J. Pritchet,
F. Valdes, D.L. Welch and an anonymous referee for their helpful
comments and/or discussions, and to thank S. Holland for his help with
the observations.

\clearpage

\begin{center}
\begin{deluxetable}{lcc}
\footnotesize
\tablecaption{Telescope coordinates of the $R$-band CCD fields.
\label{TAB1paper1}}
\tablewidth{0pt}
\tablehead{\colhead{FIELD} & \colhead{$\alpha$}  & \colhead{$\delta$}\nl
& \colhead{(2000)} & \colhead{(2000)} }
\startdata
{\sc NGC 4874}       & 12:59:35.17  & 27:57:32.43  \\
{\sc NGC 4889}       & 13:00:30.15  & 27:57:49.61  \\ 
{\sc NGC 4874 South} & 12:59:39.19  & 27:43:21.44  \\
{\sc Control}        & 13:09:03.92  & 27:59:59.42  \\ 
\enddata
\end{deluxetable}
\end{center}

\clearpage

\begin{center}
\begin{deluxetable}{lccccrccccc}
\footnotesize
\tablecaption{Properties of the master CCD images.
\label{TAB2paper1}}
\tablewidth{0pt}
\tablehead{\colhead{FIELD} & \colhead{Filter} & \colhead{Night} & 
\colhead{$\overline{X}$} & \colhead{Exp Time}  & \colhead{Master}  & 
\colhead{\#} & \colhead{$\overline{\rm FWHM}$} & \colhead{$\sigma_{\rm G}$} &
\colhead{$\overline{z_{\rm sky}}$} & \colhead{$\overline{\sigma_{\rm sky}}$}\nl
&&&& \colhead{(sec)}&& \colhead{PSF} & \colhead{PSF} && 
\colhead{({\small ADU})} & \colhead{({\small ADU})}}
\startdata
{\sc NGC 4874}    & $R$ & 1 & 1.22 & $3\times900$ & WAvg & 6 & 
2.77 & 0.52 & 1728.95 & 7.33 \\
            & $B$ & 1 & 1.09 & $3\times900$ &  Med & 6 & 3.00 &    
  &         & \\
{\sc NGC 4889}    & $R$ & 2 & 1.04 & $3\times900$ & WAvg & 7 & 
2.34 & 0.74 & 1268.96 & 6.16 \\
            & $B$ & 2 & 1.01 & $2\times900$ &  Avg & 7 & 2.93 &    
  &         & \\
{\sc NGC 4874}    & $R$ & 2 & 1.22 & $3\times900$ & WAvg & 7 & 
2.77 & 0.42 & 1457.20 & 6.40 \\
\ \ {\sc South}   & $B$ & 1 & 1.25 & $2\times900$ &  Avg & 7 & 
2.88 &      &         & \\
{\sc Control}     & $R$ & 1 & 1.02 & $3\times900$ & WAvg & 4 & 
2.92 & 0.39 & 1249.34 & 9.98 \\
            & $B$ & 1 & 1.07 & $3\times900$ &  Med & 4 & 3.06 &    
  &         & \\
\enddata
\end{deluxetable}
\end{center}

\clearpage

\begin{center}
\begin{deluxetable}{clcccc}
\footnotesize
\tablecaption{Summary of April 1991 KPNO 4m standard star fields.
\label{TAB3paper1}}
\tablewidth{0pt}
\tablehead{\colhead{NIGHT} & \colhead{Region} & \colhead{Standard} & 
\colhead{Filter} & \colhead{Exp Time} & \colhead{X} \nl
      &        &  \colhead{Stars}   &        &   \colhead{(sec)}  &}
\startdata
      1   &   M67     &   11  &   B  &   10  &  1.07 \\
          &   M67     &   11  &   R  &   ~5  &  1.07 \\
          &  NGC 2419 &   19  &   R  &   20  &  1.09 \\
          &  NGC 2419 &   10  &   B  &   50  &  1.10 \\
          &   M92     &   18  &   B  &   30  &  1.03 \\
          &   M92     &   20  &   R  &   10  &  1.03 \\
          &   M92     &   21  &   R  &   10  &  1.02 \\
          &   M92     &   20  &   B  &   30  &  1.03 \\
      2   &   M67     &   11  &   R  &   ~5  &  1.07 \\
          &   M67     &   11  &   B  &   10  &  1.07 \\
          &  NGC 4147 &   12  &   B  &   50  &  1.45 \\
          &  NGC 4147 &   16  &   R  &   20  &  1.42 \\
          &  NGC 4147 &   19  &   R  &   20  &  1.09 \\
          &  NGC 4147 &   11  &   B  &   50  &  1.12 \\
\enddata
\tablenotetext{1}{Nights 1 and 2 correspond to the nights of April 
9th and 10th, 1991.  Observations made with the TE 2K CCD at PF of 
the KPNO 4m.}
\end{deluxetable}
\end{center}

\clearpage

\figcaption[COMAfields.eps]{Three $R$-band master images
of the Coma cluster core: (a) The {\protect {\sc NGC 4874}} field, (b)
the {\protect {\sc NGC 4889}} field, and (c) the {\protect {\sc NGC
4874 South}} field.  On each image, North is towards the top and East
is to the right. Each field is $15\times18$ arcmin$^2$, and has a
significant overlap with the neighboring field. \label{COMAfields}}

\figcaption[PIXELhist.eps]{A randomly selected sample of $4\times10^4$
pixels, illustrating a typical distribution of pixel intensities.
The dotted line illustrates a Gaussian fit to the sky: the modal sky is
given by $\overline{z_{\rm sky}}$, and the standard deviation by
$\overline{\sigma_{\rm sky}}$. The {\em sky} pixels dominate in number
over the {\em object} pixels, and the sky pixels with intensities
below the mode are uncontaminated.  When sorted in intensity, the
pixels used for object detection are those above the $\beta = 4$ detection
threshold, illustrated here by the hatched line. \label{PIXELhist}}

\figcaption[RDLMNTsim.eps]{Parameter-space plots of the intensity-weighted 
radial moments $r_1$ and $r_{-2}$ versus the object's total 
input $R_T$ magnitude.  For these moments, the stellar sequence is constant
at $r_1 = 1.6$ (px) and $r_{-2}= 1.4$ (px), and the galaxies deviate
from this sequence towards larger values. \label{RDLMNTsim}}

\figcaption[FOURrdlmmnts.eps]{The intensity-weighted radial moment 
$r_{-2}$ (px)  versus the total $R_{2r_1}$ mag, for all
objects detected and measured on the four master $R$-band images: (a)
{\protect {\sc NGC 4874}}, (b) {\protect {\sc NGC 4889}}, (c)
{\protect {\sc NGC 4874 South}}, and (d) the {\protect {\sc Control}}
field.  The large number of dE galaxies in the cluster core is evident
as the excess nonstellar objects in the program fields, as compared
with the {\sc Control} field. The horizontal dotted line excludes the
bright starlike sequence and the point-like objects, while the
vertical dotted line corresponds to our adopted limiting
magnitude. \label{FOURrdlmmnts}}

\figcaption[ELLIPcomp.eps]{Difference $\Delta \epsilon$ between
the measured ($\epsilon$) and input ($\epsilon_T$) ellipticity for all
objects on the artificial CCD images, plotted against total $R$
magnitude.  For the brightest galaxies and stars, $\Delta \epsilon$ is
nearly zero and the measured values of $\epsilon$ are well
constrained.  However, for faint nonstellar objects, the measured
values of $\epsilon$ are too uncertain to prove
useful. \label{ELLIPcomp}}

\figcaption[RMAGcorrections.eps]{True (input) minus measured magnitude
for starlike and nonstellar objects on our $R$ frames.  We define the total 
magnitude $R_{2r_1}$ to include all luminosity in an aperture of
radius of $2r_1$ (px), or the radius of the seeing disk (whichever is
greater).  For faint and unresolved objects, virtually all the luminosity
is enclosed within this radius.  For the brighter objects, about 90
percent of the flux is measured for exponential-profile galaxies (open
circles) and 96 percent for the Gaussian stars (filled circles).  The
$2r_1$ total magnitude can accurately recover the true (input)
magnitude $R_T$ for uncrowded artificial objects: for galaxies, the solid
line defines the correction factor in the interval $16 \leq R_{2r_1}
\leq 24$ mag, and for stars, the dotted line illustrates the maximum
value of the correction factor. \label{RMAGcorrections}}

\figcaption[SIMcentsurf.eps]{The central surface brightness $I_{\rm c}$ 
versus the total input $R$ magnitude. At any magnitude the faint
exponential-profile galaxies have a lower central surface brightness
than do the stars. \label{SIMcentsurf}}

\figcaption[FOURcmds.eps]{Color-magnitude diagrams for all 
objects detected on the four master $R$-band images: (a) {\protect
{\sc NGC 4874}}, (b) {\protect {\sc NGC 4889}}, (c) {\protect {\sc NGC
4874 South}}, and (d) the {\protect {\sc Control}} field.  The cluster
dE galaxies are immediately obvious as the narrow near-vertical
sequence of objects not present on the {\sc Control} field. The
vertical dashed lines represent generous color limits to this dE
sequence, and the horizontal dashed line corresponds to our adopted
limiting magnitude.
\label{FOURcmds}}

\figcaption[PHOTerrors.eps]{Photometric uncertainties on our
measurements of the color (crosses) and total magnitude (solid
circles), for all objects above 22.5 mag on all four images: (a)
{\protect {\sc NGC 4874}}, (b) {\protect {\sc NGC 4889}}, (c)
{\protect {\sc NGC 4874 South}}, and (d) {\protect the {\sc Control}}
field.  Typical uncertainties are $\lesssim \pm0.06$ mag in total
magnitude, and $\lesssim \pm$0.12 mag in color.  \label{PHOTerrors}}

\end{document}